 \documentclass[preprint2]{aastex}

\slugcomment{Not to appear in Nonlearned J., 45.}

\shorttitle{Stellar kinematics in W4}
\shortauthors{Lim et al.}

\begin{document}


\title{THE ORIGIN OF A DISTRIBUTED STELLAR POPULATION 
IN THE STAR-FORMING REGION W4}


\author{Beomdu Lim \altaffilmark{1,6}, Jongsuk Hong \altaffilmark{2}, 
Hyeong-Sik Yun \altaffilmark{1}, Narae Hwang\altaffilmark{3},
Jinyoung S. Kim \altaffilmark{4}, Jeong-Eun Lee \altaffilmark{1}, 
Byeong-Gon Park \altaffilmark{3,5}, and Sunkyung Park\altaffilmark{1}}
\email{blim@khu.ac.kr}

\altaffiltext{1}{School of Space Research, Kyung Hee University 1732, Deogyeong-daero, Giheung-gu, Yongin-si, Gyeonggi-do 17104, Republic of Korea}
\altaffiltext{2}{Department of Astronomy, Yonsei University 50, Yonsei-ro, Seodaemun-gu, Seoul 03722, Republic of Korea}
\altaffiltext{3}{Korea Astronomy and Space Science Institute, 776 Daedeokdae-ro, Yuseong-gu, Daejeon 34055, Korea}
\altaffiltext{4}{Steward Observatory, University of Arizona, 933 N. Cherry Ave. Tucson, AZ 85721-0065, USA}
\altaffiltext{5}{Astronomy and Space Science Major, University of Science and Technology, 217 Gajeong-ro, Yuseong-gu, Daejeon 34113, Republic of Korea}

%\and


\altaffiltext{6}{Corresponding author}


\begin{abstract}
Stellar kinematics provides the key to understanding the formation process and 
dynamical evolution of stellar systems. Here, we present a kinematic study of the 
massive star-forming region W4 in the Cassiopeia OB6 association using the {\it Gaia} 
Data Release 2 and high-resolution optical spectra. This star-forming region is 
composed of a core cluster (IC 1805) and a stellar population distributed over 20 
pc, which is a typical structural feature found in many OB associations. 
According to a classical model, this structural feature can be understood in the 
context of the dynamical evolution of a star cluster. The core-extended structure 
exhibits internally different kinematic properties. Stars in the core have an almost 
isotropic motion, and they appear to reach virial equilibrium given their 
velocity dispersion ($0.9 \pm 0.3$ km s$^{-1}$) comparable to that in a virial 
state ($\sim$0.8 km s$^{-1}$). On the other hand, the distributed population 
shows a clear pattern of radial expansion. From the $N$-body simulation for the 
dynamical evolution of a model cluster in subvirial state, we reproduce the observed 
structure and kinematics of stars. This model cluster experiences collapse for the first 2 Myr. 
Some members begin to radially escape from the cluster after the initial collapse, 
eventually forming a distributed population. The internal structure and kinematics 
of the model cluster appear similar to those of W4. Our results support the idea that 
the stellar population distributed over 20 pc in W4 originate from the dynamical 
evolution of IC 1805.
\end{abstract}


\section{INTRODUCTION}
OB associations are prominent stellar systems in galaxies \citep{RW93,BKS96,
PGFP01,GHV09} because of the high-mass star population spread over a few 
tens of parsecs \citep{A47}. Massive stars in such systems are engines that 
generate giant H {\scriptsize \textsc{II}} bubbles and play a decisive role in the 
chemical evolution of host galaxies. In addition, the majority of stars form in 
associations or clusters within them \citep{LL03}, and thereby field stars 
in the Galactic disk are considered to originate from the dissolution of these 
stellar systems \citep{MS78,BPS07}. Despite their significant contribution to 
stellar populations in host galaxies, our understanding of their formation 
and evolution is still incomplete.

These large stellar systems are, in general, composed of a single or 
multiple star clusters in the central region and a distributed stellar population 
(hereafter DSP) at a large spatial scale  \citep{B64,KLB12}. These structural 
features may contain the clue to the formation process of these stellar 
systems. A classical model suggests that embedded clusters could undergo 
expansion after rapid gas expulsion \citep{LMD84,KAH01}. This dynamical 
process results in scattering of cluster members, and eventually leads to 
the formation of unbound OB associations. In addition, the structure 
of young star clusters can be highly affected by stellar feedback as it impacts 
on the timescale of gas expulsion and the dynamics of star clusters just before 
gas evacuation events \citep{GBRT17}.
 
Some star clusters in several OB associations exhibit a sign of mass 
segregation \citep{HH98,CDZ07,SSB13}. Dynamical mass segregation 
occurs on a timescale comparable to the relaxation time of given stellar 
systems, and it takes longer than ten crossing times for clusters with 
about 1000 members \citep{BD98}. However, clusters in OB associations 
are mostly younger than their relaxation times \citep{MJD95,HH98,PS02,SCB00}. 
The stellar velocity dispersions measured in the Orion Nebula 
Cluster, NGC 2244, and NGC 6530 are weakly correlated with 
stellar masses \citep{JW88,CDZ07}. Hence, it has been claimed 
that the observed internal structures could have been formed by 
star formation {\it in situ} \citep{BBZ98}, rather than dynamical 
evolution via energy equipartition. 

On the other hand, there are some attempts to understand the 
origin of mass segregation in terms of early dynamical evolution 
\citep{MVP07,AGP09,AGP10}. These theoretical studies considered 
the situation where several subgroups of stars in subvirial state form 
along the substructures in molecular clouds. Merging of these 
subgroups results in mass-segregated star clusters on a short 
timescale. Hence, it is still necessary to investigate the dynamics of 
stellar associations to understand their formation process.

In this study, we report the signature of early 
dynamical evolution in the massive star-forming region (SFR) 
W4 within the Cassiopeia OB6 association. W4 hosts a large 
number of massive OB stars and low-mass young stellar 
objects \citep{WSR10,SBC17,RZT19}. This young stellar population 
forms a single central cluster (IC 1805) and a large structure that 
extends over 20 pc \citep{SBC17}. This simple structural 
feature, compared to the other associations, provides us a better 
chance to isolate the history of the dynamical evolution in massive 
SFR. The observation and data that we used are described in 
Section 2. The selection of the kinematic members are addressed 
in Section 3. We probe the motions of stars using the {\it Gaia} proper 
motions (PMs) \citep{gaia18} and radial velocities (RVs) in Section 4. 
The dynamical state of this SFR is also investigated from velocity 
dispersions. From comparison with the results of an N-body 
simulation, a plausible explanation for the formation process 
of W4 is suggested in Section 5. Finally, we summarize our 
results in Section 6.

\section{Data}
Member candidates were obtained from the photometric 
catalogue of \citet{SBC17}. We selected stars with either `E' or `e' 
flag in this catalogue as early-type (O- or B-type) star candidates. 
On the other hand, stars satisfying at least one of the following criteria: 

\begin{enumerate}
\item H$\alpha$ emission stars or candidates
\item X-ray emission stars or candidates
\item young stellar objects showing a flat spectrum
\item Class II young stellar objects
\item young stellar objects with transition disks or pre-transition disks
\item Class III young stellar objects in PMS locus (see figure 22 of \citealt{SBC17})
\end{enumerate}

\noindent are considered as pre-main sequence (PMS) star candidates. 
The RVs of these PMS star candidates have not yet been measured.
Since PMS stars down to $\sim1$ $M_{\sun}$ in W4 can be observed 
with large telescopes, we selected PMS star candidates brighter 
than 19 mag in visual band (or $G \lesssim 18$ mag) for spectroscopic 
observation. The number of targets in our sample is 358 (115 early-type 
star and 243 PMS star candidates) in total. 

\subsection{Radial velocities}
The optical spectra of low-mass stars contain a large number of 
metallic absorption lines. In addition, the rotational velocities of these 
low-mass young stars are smaller than those of high mass stars. Therefore, 
we can more precisely measure their RVs from high-resolution spectra 
compared to high mass stars. We observed 198 low-mass PMS star 
candidates on 2018 November 29 using the high-resolution 
($R \sim 34000$) multi-object spectrograph {\it Hectochelle} 
\citep{SFC11} on the 6.5-m telescope of the MMT observatory. The 
spectra of these stars were taken with the RV31 filter in $2\times2$ 
binning mode to achieve good signal-to-noise ratios. Dome flat and 
ThAr lamp spectra were also obtained just before and after the target 
observations. 

All mosaiced frames were merged into single frames after 
overscan correction using the IRAF/{\tt MSCRED} packages. 
We extracted one-dimensional (1D) spectra from the merged 
frames using the {\tt dofiber} task in the IRAF/{\tt SPECRED} 
package. Dome flat spectra were used to correct for the 
pixel-to-pixel variation. The solutions for the wavelength calibration 
obtained from ThAr spectra were applied to the target spectra. 
We obtained a master sky spectrum with an improved 
signal-to-noise ratio by median combining the spectra from 
a few tens of fibers assigned to the blank sky for a given setup. 
Target spectra were subtracted by an associated master sky 
spectrum and then combined into a single spectrum for the 
same target. Finally, target spectra were normalized by using 
continuum levels.
 
\begin{figure*}[t]
\includegraphics[width=5.3cm]{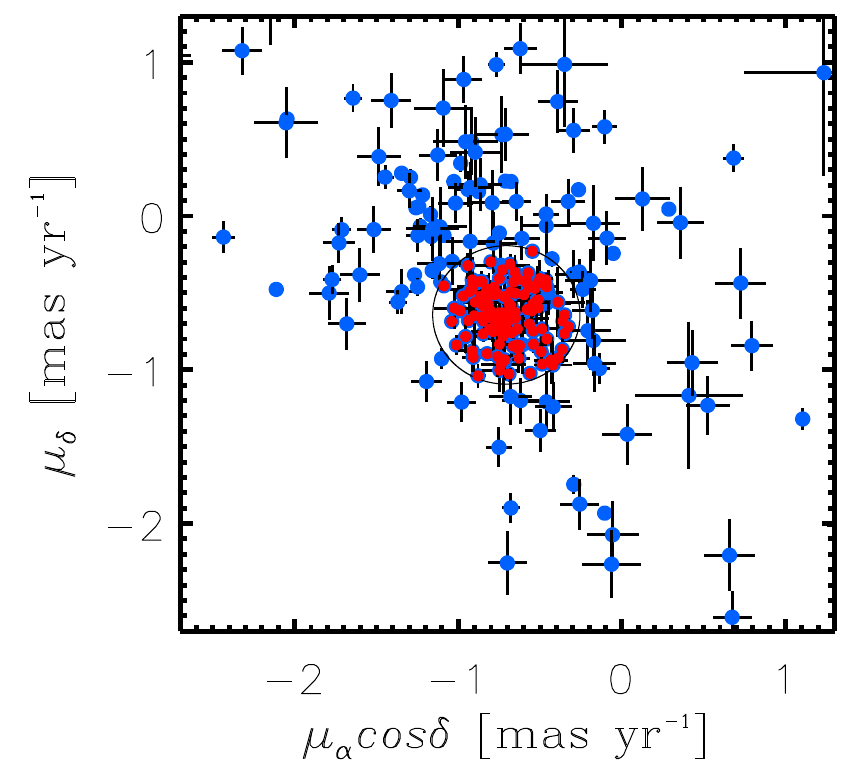}\includegraphics[width=5.3cm]{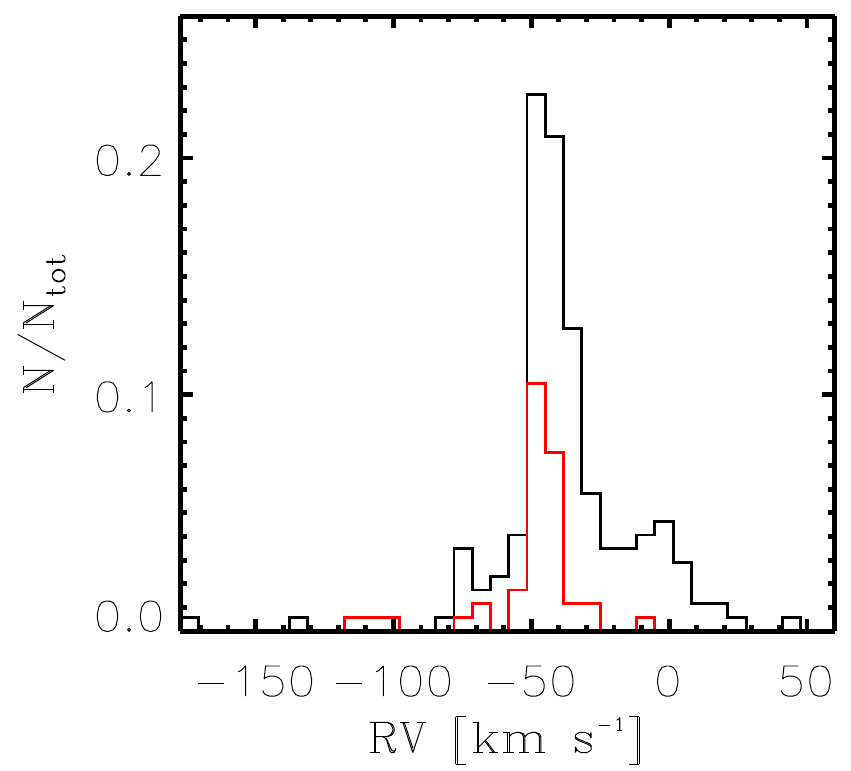}\includegraphics[width=5.3cm]{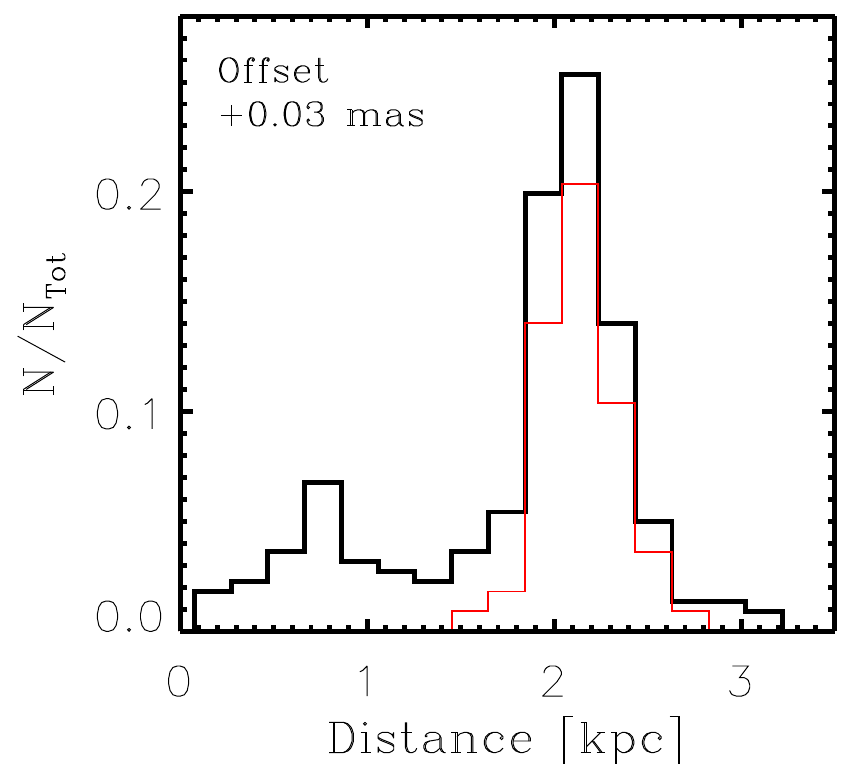}
\caption{Distributions of PMs ({\it left}), RVs ({\it middle}), 
and distances ({\it right}). The PMs and parallaxes were obtained from 
the {\it Gaia} DR2 \citep{gaia18}, while the RVs were 
measured in this study. In the left panel, blue and red symbols are the 
member candidates from \citet{SBC17} and the genuine members, respectively. The 
distributions of the member candidates are plotted by black 
histograms in the middle and right panels, while red histograms 
represent those of the genuine members. In the right panel, a systematic 
zero point offset of 0.03 from \citet{LHB18} was applied to the {\it Gaia} 
parallaxes, and then distances were computed by the inversion of the 
corrected parallaxes. Some stars with parallaxes five times smaller 
than the associated errors were excluded. The size of bins was determined 
using the relation of \citet{FD81}. 
 }\label{fig1}
\end{figure*}

Our optical spectra contain a number of metallic absorption 
lines between 5150--5300 \AA. To measure the RVs of the 
PMS stars, we applied a cross-correlation technique to the 
observed spectra. Several synthetic spectra adopting the 
Solar abundance were generated in the wide temperature 
range of 3800--9880 K using the \textsc{MOOG} code and 
Kurucz ODFNEW model atmosphere \citep{S73,CK04}. These 
spectra were used as template spectra. We derived cross-correlation 
functions (CCF) of the observed spectra using the {\tt xcsao} task in 
the \textsc{RVSAO} package \citep{KM98} and selected the synthetic 
spectra that have the strongest CCF peak values. RVs were 
measured from the derived CCF peaks. The errors on RVs were 
estimated from the relation expressed as $3w/8(1+h/\sqrt{2}\sigma_a)$, 
where $w$, $h$, and $\sigma_a$ represent the full widths at 
half-maximum of CCFs, their amplitudes, and the rms from 
antisymmetric components, respectively \citep{TD79,KM98}. 
We measured the RVs of 172 out of 243 PMS star candidates 
in total. These RVs were converted to velocities in the local 
standard of rest frame using the \textsc{IRAF}/{\tt RVCOR} task. 
The spectra of the other 26 stars have insufficient signals to derive CCFs. 

\subsection{{\it Gaia} data}
The parallaxes and PMs of the member candidates were obtained 
from the {\it Gaia} Data Release 2 (DR2) \citep{gaia18}. Their 
counterparts were searched for in this catalogue with a 
searching radius of $1\farcs0$. All but one have 
counterparts in the {\it Gaia} DR2. Errors on parallax and PMs 
from the {\it Gaia} catalogue \citep{gaia18} were adopted. These errors 
are correlated with the brightness of stars. The errors 
on parallax are about 0.03 mas on average for 
stars brighter than $G \sim 13$ mag and better than 
0.20 mas for stars brighter than $G \sim 18$ mag. The 
mean values of PM errors along the R.A. and declination 
are about 0.04 and 0.05 mas yr$^{-1}$, respectively, and 
these errors increase up to 0.20--0.25 mas yr$^{-1}$ 
for faint stars.

We did not use stars with negative parallaxes or close companion 
(duplication flag = 1), or without astrometric parameters in analysis. 
The distances to individual stars were computed by the inversion 
of the parallaxes \citep{gaia18} after correction for a systematic 
zero point of 0.03 mas from \citet{LHB18}.

\section{MEMBERSHIP VALIDATION}
In \citet{SBC17}, the member candidates were selected using 
multi-color photometric diagrams and a list of X-ray sources 
\citep{RN16}. However, their selection criteria are insufficient 
to isolate genuine members because a number of objects 
with similar properties to young stars can scatter along the 
line of sight. The most reliable members can be selected 
when combined with astrometric parameters.

Fig.~\ref{fig1} displays the distributions of PMs, RVs, and 
distances for the member candidates. The most probable 
members form a dense group in the PM plane, while some 
candidates are distributed over a wide range of PMs. Similarly, 
the RV distribution has a strong peak at $-45$ km s$^{-1}$ 
with an extended wing component. In the distance distribution, 
there are foreground stars closer than 1.3 kpc. Thus, the 
candidates in the extended components are nonmembers.

In order to select genuine members, stars that are either 
closer than 1.3 kpc or farther than 3.0 kpc were first 
excluded. We set a circular region around a median PM 
value to encompass probable members (see the left panel 
in Fig.~\ref{fig1}). In this region, stars with PMs within three 
times the standard deviations from the weighted mean values were selected 
as members, where the inverse of the squared PM error 
was used as the weight value. We obtained a new 
median PM value from the PMs of stars satisfying this 
criterion, and this procedure was repeated three times.
A total of 127 members (55 early-type and 72 PMS stars) 
were selected. RVs were measured for about 62\% 
of these PMS members (45/72). The RVs of the 
PMS members show a single Gaussian distribution 
(see the middle panel of Fig.~\ref{fig1}). 

Their distance distribution also has a single Gaussian 
peak at 2.1 kpc with a width of 0.2 kpc (the 
right panel of Fig.~\ref{fig1}). We adopted this value as 
the distance to W4. Systematic errors of $\pm$0.2 kpc 
may exist due to the reported zero point offsets 
\citep{LHB18,ST18,ZPHS19}, however, this distance is 
reasonably consistent with previous 
studies within errors \citep{JS61,H78,MJD95,SBC17,CG18}. 
We present a list of the kinematic members and their 
data in Table.~\ref{tab1}.

\citet{CG18} also selected the member candidates of IC 1805 
using the {\it Gaia} DR2 \citep{gaia18} and published 
their catalogue. The number of members with membership 
probabilities greater than 0.5 is 136 in total, of which 83 
overlap with our members. However, some member 
candidates in their catalogue are found below the PMS 
locus in the ($G$, $B_{\mathrm{p}} - R_{\mathrm{p}}$) 
color-magnitude diagram. In addition, there are a few member 
candidates with far different PMs from the mean PM of IC 1805. 
We thus suggest to use member candidates with higher 
membership probabilities.

\begin{center}
\begin{deluxetable}{lccccccccccccccccc}
\rotate 
\setlength\tabcolsep{1.5pt}
\tabletypesize{\tiny}
\tablewidth{0pt}
\tablecaption{Catalogue of kinematic members. \label{tab1}}
\tablehead{\colhead{Sq.} & \colhead{R.A. (2000)} & \colhead{Dec. (2000)} & \colhead{$\pi$} & \colhead{$\epsilon(\pi)$} &
\colhead{$\mu_{\alpha}\cos\delta$}  & \colhead{$\epsilon(\mu_{\alpha}\cos\delta)$} &
\colhead{$\mu_{\delta}$} & \colhead{$\epsilon(\mu_{\delta})$} & \colhead{$G$} & \colhead{$\epsilon(G)$} &
\colhead{$B_P$} & \colhead{$\epsilon(B_P)$} & \colhead{$R_P$} & \colhead{$\epsilon(R_P)$} & \colhead{RV} & \colhead{$\epsilon$(RV)} &
\colhead{Binarity flag} \\
 & \colhead{(h:m:s)} & \colhead{($\degr:\arcmin:\arcsec$)} &  \colhead{(mas)} & \colhead{(mas)} & \colhead{(mas yr$^{-1}$)} &\colhead{(mas yr$^{-1}$)} &
\colhead{(mas yr$^{-1}$)} &\colhead{(mas yr$^{-1}$)} & \colhead(mag) & \colhead(mag) & \colhead(mag) & \colhead(mag) & \colhead(mag) & \colhead(mag) &
\colhead{(km s$^{-1}$)} & \colhead{(km s$^{-1}$)} &  }
\startdata
   1  & 02:29:45.93 & +61:34:42.1 &  0.3577 &  0.0587 &  -0.531 &  0.062 &  -0.471 &  0.097 & 16.3977 & 0.0005 & 17.4011 & 0.0056 & 15.3683 & 0.0016 &  -43.785 &  0.445  &     \\ 
  2  & 02:30:12.25 & +61:24:45.7 &  0.3473 &  0.0848 &  -0.914 &  0.085 &  -0.877 &  0.136 & 16.7894 & 0.0014 & 18.0058 & 0.0094 & 15.6706 & 0.0055 & -110.530 &  1.163  &     \\
  3  & 02:30:15.39 & +61:23:42.1 &  0.4242 &  0.0206 &  -1.040 &  0.022 &  -0.685 &  0.033 & 13.2875 & 0.0002 & 13.9044 & 0.0012 & 12.5116 & 0.0008 &  \nodata & \nodata &     \\
  4  & 02:30:57.84 & +61:14:26.7 &  0.3402 &  0.1175 &  -0.902 &  0.128 &  -0.648 &  0.203 & 17.7242 & 0.0023 & 18.8353 & 0.0315 & 16.6204 & 0.0095 &  -44.269 &  3.120  &     \\
  5  & 02:31:11.81 & +61:43:38.4 &  0.3308 &  0.0412 &  -1.011 &  0.042 &  -0.839 &  0.068 & 13.6276 & 0.0004 & 13.9894 & 0.0011 & 13.0748 & 0.0016 &  \nodata & \nodata &     \\
  6  & 02:31:13.95 & +61:30:46.0 &  0.4859 &  0.1148 &  -0.807 &  0.144 &  -0.610 &  0.194 & 17.7973 & 0.0019 & 18.9303 & 0.0152 & 16.6720 & 0.0058 &  \nodata &  0.000  &     \\
  7  & 02:31:37.48 & +61:28:13.9 &  0.4994 &  0.1216 &  -0.913 &  0.138 &  -0.476 &  0.211 & 17.9111 & 0.0017 & 19.0084 & 0.0182 & 16.7412 & 0.0077 &  -28.168 &  4.415  &    \\ 
  8  & 02:31:47.85 & +61:27:32.5 &  0.4902 &  0.0279 &  -0.897 &  0.031 &  -0.560 &  0.050 & 15.0796 & 0.0003 & 15.6026 & 0.0019 & 14.3635 & 0.0011 &  -10.097 &  1.564  &    \\ 
  9  & 02:31:48.48 & +61:34:56.0 &  0.4104 &  0.0176 &  -0.869 &  0.020 &  -0.427 &  0.031 & 13.6410 & 0.0006 & 14.1459 & 0.0026 & 12.9615 & 0.0024 &  \nodata & \nodata &    \\ 
 10  & 02:31:49.79 & +61:32:41.4 &  0.4580 &  0.0167 &  -0.865 &  0.020 &  -0.523 &  0.032 & 13.2861 & 0.0002 & 13.6591 & 0.0014 & 12.7244 & 0.0008 &  \nodata & \nodata &    \\ 
 11  & 02:31:50.22 & +61:35:59.9 &  0.4067 &  0.0369 &  -0.816 &  0.039 &  -0.595 &  0.066 & 15.6844 & 0.0007 & 16.6313 & 0.0036 & 14.6993 & 0.0025 &  -72.868 &  0.714  &    \\ 
 12  & 02:31:50.73 & +61:33:32.3 &  0.4089 &  0.0388 &  -0.683 &  0.045 &  -0.313 &  0.067 & 15.9216 & 0.0006 & 16.8135 & 0.0051 & 14.9518 & 0.0032 &  -38.942 &  7.406  &    \\ 
 13  & 02:31:55.15 & +61:31:22.9 &  0.3927 &  0.0756 &  -0.915 &  0.095 &  -0.458 &  0.130 & 17.0618 & 0.0014 & 18.1185 & 0.0100 & 15.9640 & 0.0044 &  -46.485 &  6.412  &    \\ 
 14  & 02:32:02.43 & +61:37:13.2 &  0.4941 &  0.0297 &  -0.827 &  0.034 &  -0.491 &  0.052 & 15.2706 & 0.0003 & 15.9645 & 0.0016 & 14.4432 & 0.0011 &   \nodata&  0.000  &    \\ 
 15  & 02:32:06.48 & +61:29:54.3 &  0.4419 &  0.0178 &  -0.846 &  0.020 &  -0.683 &  0.031 & 13.5996 & 0.0002 & 14.1713 & 0.0014 & 12.8521 & 0.0007 &  \nodata & \nodata &    \\ 
 16  & 02:32:07.03 & +61:45:33.8 &  0.4831 &  0.0467 &  -0.941 &  0.052 &  -0.322 &  0.085 & 15.8931 & 0.0010 & 16.9697 & 0.0067 & 14.8460 & 0.0029 &   \nodata&  0.000  &    \\ 
 17  & 02:32:07.91 & +61:24:51.4 &  0.4220 &  0.0352 &  -0.736 &  0.042 &  -0.962 &  0.062 & 15.4726 & 0.0040 & 16.3413 & 0.0149 & 14.4963 & 0.0075 &  -57.268 &  1.667  &    \\ 
 18  & 02:32:09.63 & +61:38:23.5 &  0.4338 &  0.0277 &  -0.914 &  0.033 &  -0.415 &  0.047 & 11.3079 & 0.0042 & 11.5875 & 0.0086 & 10.8703 & 0.0129 &  \nodata & \nodata &    \\ 
\enddata
\tablecomments{Column (1) : Sequential number. Columns (2) and (3) : The equatorial coordinates of members. Columns (4) and (5) : Absolute parallax and its
standard error. Columns (6) and (7) : PM in the direction of right ascension and its standard error. Columns (8) and (9): PM in the direction
of declination and its standard error. Columns (10) and (11) : $G$ magnitude and its standard error. Columns (12) and (13) : $B_P$ magnitude and
its standard error. Columns (14) and (15) : $R_P$ magnitude and its standard error. Columns (16) and (17) : RV and its error. Columns (18) : Binarity flag. SB2 represents 
a double-lined spectroscopic binary candidate. The parallax and PM were taken from the Gaia Data Release 2 \citep{gaia18}. The full table is available electronically.}
\end{deluxetable}
\end{center}

\begin{figure}[!]
\includegraphics[width=7.3cm]{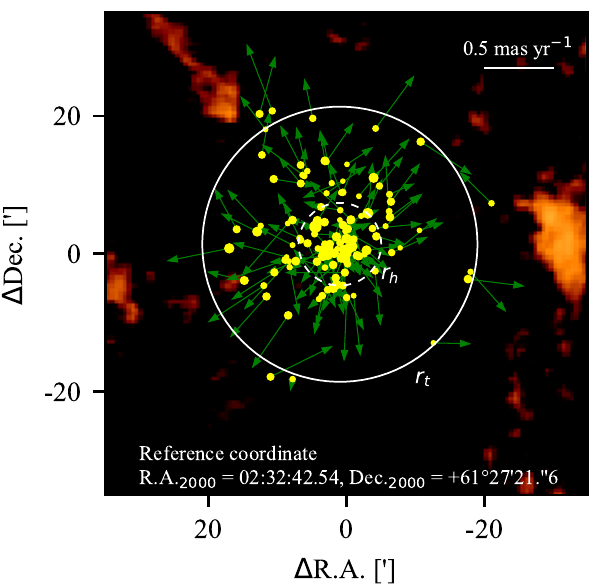}
\includegraphics[width=7.3cm]{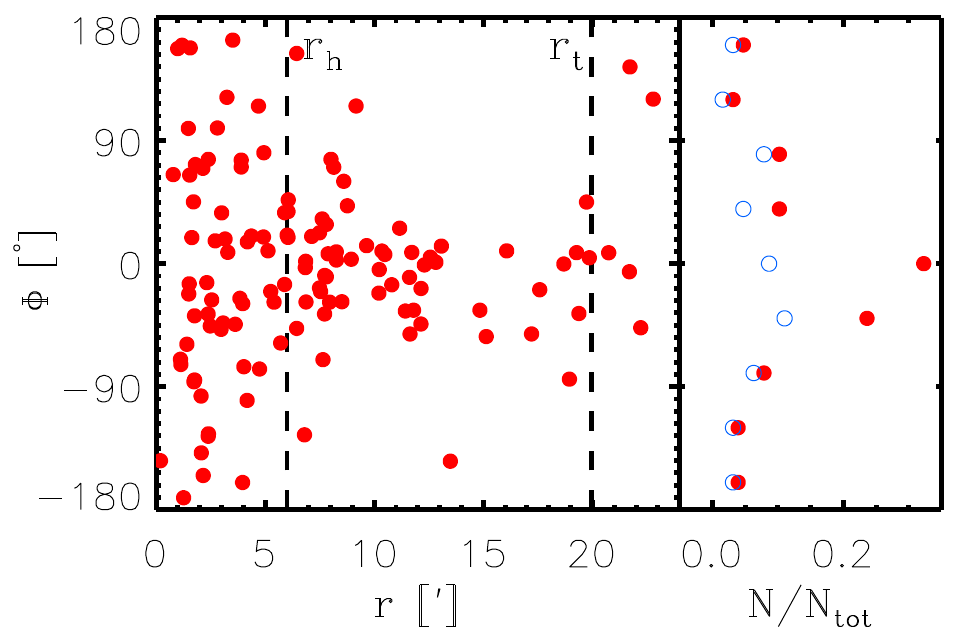}
\caption{Relative motions of the members in W4. In the 
upper panel, the spatial distribution of members are 
overplotted on the zeroth moment map of the $^{12}$CO 
($J = 1 - 0$) line from \citet{HBS98}. The size of yellow dots is 
proportional to the brightness of individual stars, and 
green arrows denote relative PM vectors. The two white 
circles (dashed and solid lines) represent $r_h$ and $r_t$, 
respectively. In the lower left panel, the angles ($\Phi$) 
between the relative PM and radial vector (from the cluster 
center to a given star) are plotted with respect to the 
central distances. Each dashed line represents $r_h$ and $r_t$. 
The lower right panel displays the number distributions of 
stars with different $\Phi$ values from different sample; all 
members (red filled circle) and members within $r_h$ (blue open 
circle). These numbers were normalized by the total number 
of the kinematic members.  }
\label{fig2}
\end{figure}

\section{DYNAMICAL STATE OF W4}
Fig.~\ref{fig2} displays the spatial distribution of 
the selected members. While a number of stars are found 
in the central region, some stars form a large structure 
that extends out to about 40$\arcmin$ (equivalent to 24 pc). 
This extended structure is not featured by inclusion of 
nonmembers. A half-number radius ($r_h$) of this SFR 
obtained from our sample is about of 6$\arcmin$ 
(3.7 pc), which is consistent with that ($6\farcm7$) 
determined by \citet{CG18}. We define the structure 
within $r_h$ as the core and that outside this radius 
as the DSP.

We computed the tidal radius ($r_t$) of this SFR 
using the equation 3 of \citet{K62}. The total masses of 
the cluster and the Galaxy were taken from previous 
studies (2700$M_{\sun}$ from \citealt{SBC17} and $1.3 \times 
10^{12}M_{\sun}$ from \citealt{M17}, respectively). The tidal radius is 
estimated to be about 12 pc. The majority of the members 
exist within the tidal radius, which implies that the origin 
of the DSP is not the tidal disruption by the Galaxy. 

The median PMs of the members are $-0.706$ mas yr$^{-1}$ 
and $-0.643$ mas yr$^{-1}$ along the R.A. and declination, 
respectively. These values are in good agreement with the 
mean PMs ($-0.702$ mas yr$^{-1}$, $-0.669$ mas yr$^{-1}$) 
obtained by \citet{CG18}. The PM vectors of the members relative 
to the median PMs show outward motions (Fig.~\ref{fig2}). This is the 
typical pattern of expansion as seen in other OB associations 
\citep{KHS19,LNGR19}. To investigate these motions in detail, 
we computed the angle ($\Phi$) between the radial vector 
of a given star from the cluster center (median coordinates) 
and its relative PM vector as introduced in our previous 
paper \citep{LNGR19}. Note that a $\Phi$ value of 0$^{\circ}$ 
indicates radial expansion. Members in the core exhibit 
a uniform $\Phi$ distribution (see the lower panels of 
Fig.~\ref{fig2}), which indicates that the directions of their 
motions are almost isotropic. On the other hand, the 
members belonging to the DSP clearly show a radial 
expansion. Thus, the DSP seems to be a group of stars 
radially escaping from the core. These results are 
very similar to the kinematic properties of escaping 
stars from the Orion Nebula Cluster \citep{PRB20}.

\begin{figure}[t]
\epsscale{1.0}
\plotone{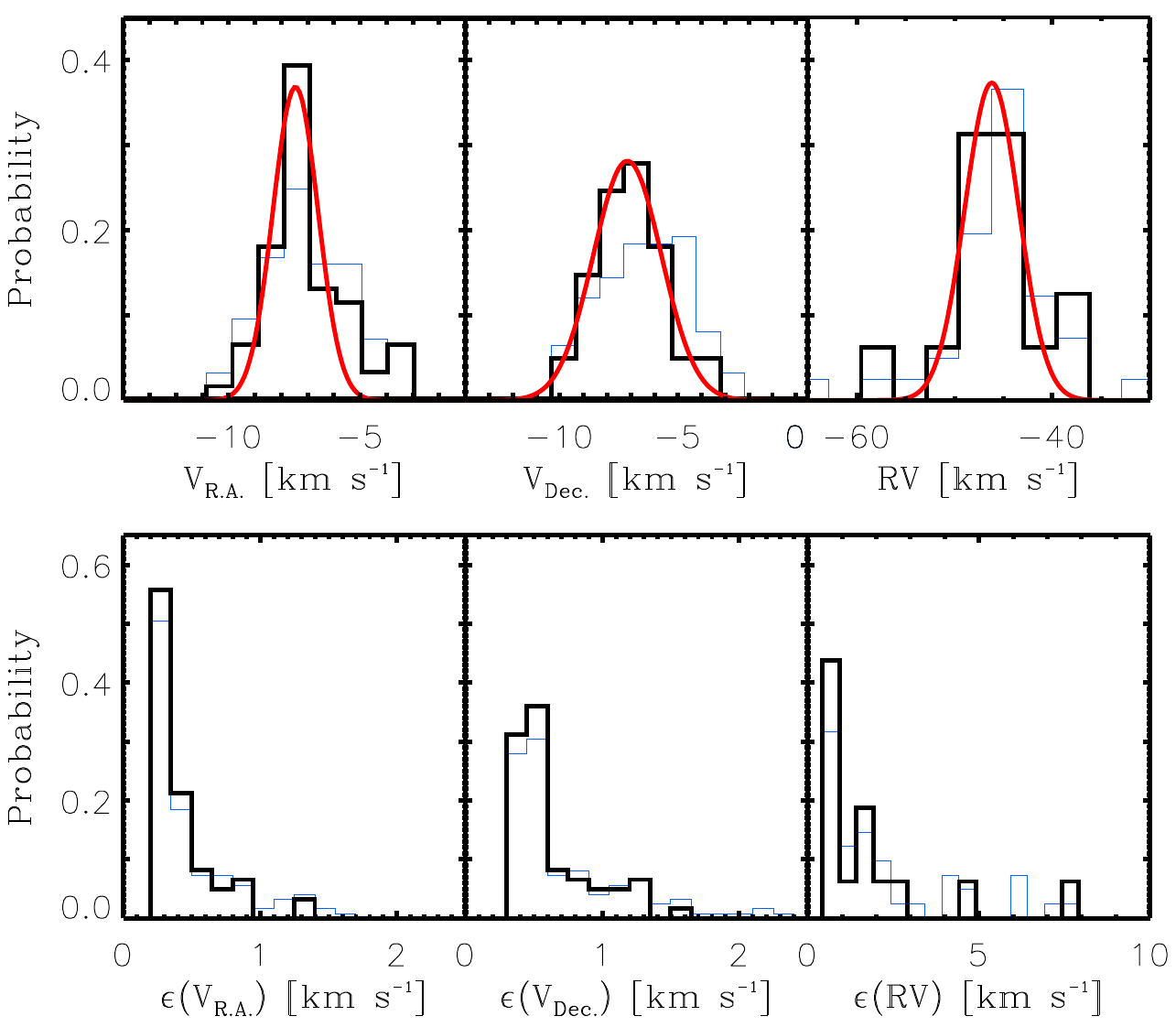}
\caption{Velocity (upper) and error (lower) distributions 
along R.A., declination, and the line of sight. Blue (thin) and 
black (thick) histograms were obtained from all members 
and only the members within $r_h$, respectively. Upper 
panels: Bin sizes were determined in the same manner as adopted in 
Fig.~\ref{fig1}. The best-fit Gaussian distributions for the 
members within $r_h$ are plotted by red solid lines. Lower 
panels: Bin sizes are 0.15, 0.15, and 0.50 km s$^{-1}$ from left 
to right. All the histograms were normalized by the total 
number of stars. }\label{fig3}
\end{figure}

We investigate the dynamical state of this SFR using 
the velocity dispersions among the members. The 1D 
velocities along R.A. and declination were calculated 
multiplying PMs by the distance of 2.1 kpc. The errors 
due to the differences of distances among individual 
members are expected to be less than 1\% because 
the extent of internal structure along the line of sight 
is very small compared to the distance to W4 assuming 
spherical symmetry. For the spectroscopic sample, 
stars in the RV range of $-75$ km s $^{-1}$ to $-15$ km s$^{-1}$ 
were used to minimize the contributions of close 
binary candidates with large amplitudes. 

Fig.~\ref{fig3} displays the distributions of 1D velocities along R.A., 
declination, and the line of sight ($V_{\mathrm{R.A.}}, V_{\mathrm{Dec}}$, 
and RV). All the distributions appear close to the Gaussian distribution. 
Velocity dispersions were measured from the best-fit Gaussian 
widths. However, since the DSP is considered to be a group 
of stars escaping from the central cluster, including these 
stars can overestimate the kinematic velocity dispersion of 
this SFR. A similar aspect was found in the central region of 
the Orion Nebula Cluster \citep{KLK19}. We therefore used 
only the members in the core. Indeed, the velocity dispersions 
for all the members appear larger than those for stars in the core 
(see Fig.~\ref{fig3} and Table~\ref{tab2}). 

\begin{table*}
\begin{center}
\setlength\tabcolsep{5pt}
\caption{Velocity dispersions. \label{tab2}}
\begin{tabular}{lccccccccc}
\tableline\tableline\scriptsize
Sample & $\sigma_{\mathrm{R.A.}}$ & $\epsilon_{\mathrm{obs,R.A.}}$ & $\sigma_{\mathrm{int,R.A.}}$ & $\sigma_{\mathrm{Dec.}}$ & $\epsilon_{\mathrm{obs,Dec.}}$ & $\sigma_{\mathrm{int,Dec.}}$ & $\sigma_{\mathrm{RV}}$ & $\epsilon_{\mathrm{obs,RV}}$ & $\sigma_{\mathrm{int,RV}}$ \\
 & (km s$^{-1}$) & (km s$^{-1}$) & (km s$^{-1}$) & (km s$^{-1}$) & (km s$^{-1}$) & (km s$^{-1}$) & (km s$^{-1}$) & (km s$^{-1}$) & (km s$^{-1}$)\\
\tableline
All & 1.88 & 0.50 & 1.81 & 2.21 & 0.74 & 2.08 & 2.78 & 2.64 & 0.87 \\
$r< r_h$ & 0.86 & 0.43 & 0.74 & 1.44 & 0.74 &  1.24 & 2.76 &  2.64 & 0.79\\ 
\tableline
\end{tabular}
\tablecomments{Velocity dispersions were obtained for all members and members 
within $r_h$, respectively. Columns (2), (5), and (8) represent the measured 
velocity dispersions along R.A, Dec., and RV, respectively. The errors corresponding 
to each measurement are shown in columns (3), (6), and (9). Columns (4), (7), and 
(10) denote intrinsic velocity dispersions.}
\end{center}
\end{table*}

The representative observational errors were estimated from the 
weighted-mean of errors; where the weights were adopted from 
the probability functions presented in the lower panels of 
Fig.~\ref{fig3}. The intrinsic velocity dispersions of
$\sigma_{\mathrm{int,RA}}$, $\sigma_{\mathrm{int,Dec}}$, 
and $\sigma_{\mathrm{int,RV}}$ were then calculated to 
be about 0.74, 1.24, and 0.79 km s$^{-1}$, respectively, after 
quadratic subtraction of the typical errors from the measured 
velocity dispersions. Systematic errors of $\pm$0.1 km s$^{-1}$ 
due to the zero point offsets in parallax can be considered 
for $\sigma_{\mathrm{int,RA}}$ and $\sigma_{\mathrm{int,Dec}}$. 
We adopted the mean value of these intrinsic velocity dispersions 
[$0.9 \pm 0.3$ (s.d.) km s$^{-1}$] as the 1D velocity dispersion of W4.

The 1D velocity dispersion in virial state is given 
by the following equation \citep{PW16};

\begin{equation}
\sigma_{\mathrm{vir}} = \sqrt{{2GM_{\mathrm{total}} \over \eta R}}
\end{equation}

\noindent where $G$, $M_{\mathrm{total}}$, $R$, and $\eta$ 
represent the gravitational constant, enclosed mass, radius, and 
the structure parameter, respectively. This SFR contains little 
gas inside the H {\scriptsize \textsc{II}} bubble according to the zeroth 
moment map of $^{12}$CO $J= 1-0$ taken from \citet{HBS98} 
(see Fig.~\ref{fig2}). Therefore, the total enclosed mass was assumed 
to be the total stellar mass of 2700$M_{\sun}$ derived by \citet{SBC17}. 
The $r_h$ of 3.7 pc was adopted in equation 1. For $\eta$, star clusters 
have a value between 1 to 11 depending on their surface densities 
\citep{PMG10}. W4 has a surface density profile with a core radius 
($r_c$) of 0.7 pc \citep{SBC17}. Since the concentration parameter 
$\log r_t / r_c$ ($\sim$ 1.2) is smaller than 1.8, the $\eta$ value 
of 9.75 was adopted \citep{PMG10}. The virial velocity dispersion 
of this SFR was then estimated to be about 0.8 km s$^{-1}$. The 
error on total stellar mass ($\pm$300 $M_{\sun}$) from \citet{SBC17} 
does not significantly affect the resultant velocity dispersion (less than 
$\pm$0.1 km s$^{-1}$). The virial velocity dispersion is comparable to the 
measured one within the observational error. Therefore, our result 
indicates that the motions of stars in the core are close to virial 
equilibrium. 

The adopted distance (2.1 kpc) is slightly smaller than that (2.4 kpc) 
derived by \citet{SBC17}. To test the effect of different distances 
on the total stellar mass, we simulated a simple stellar population 
with an age of 3.5 Myr using a Monte-Carlo technique. A total of 4500 
stars were generated, based on the initial mass function of \citet{K01}. 
Its total stellar mass is about 2771$M_{\sun}$. The bolometric 
magnitudes and effective temperatures of these stars were obtained 
by interpolating their masses to evolutionary models for main sequence 
\citep{EGE12} and PMS stars \citep{SDF00}. We then dimmed their bolometric 
magnitudes by 0.3 mag. The masses of the artificial stars were rederived 
by interpolating the bolometric magnitudes and effective temperatures 
to the evolutionary tracks \citep{SDF00,EGE12}. A total stellar 
mass was derived from the sum of these masses. As a result, 
the difference of 0.3 mag in distance modulus resulted in only about 
5\% error in total stellar mass.

\section{THE ORIGIN OF THE DISTRIBUTED STELLAR POPULATION}
Using the NBODY6++GPU code \citep{WSA15}, we conducted 
the $N$-body simulation of a model cluster to understand 
the observed structural and kinematic features in the context 
of dynamical evolution. The initial number of stars was set to 5,000, and 
their masses were drawn from the Kroupa initial mass 
function in the range from 0.1 to 100 $M_{\sun}$ \citep{K01}. 
We adopted the density profile of \citet{K66} with the 
dimensionless concentration parameter $W_0 = 3$. The initial 
half-mass radius and cutoff radius were set to be 2.3 pc and 
8.6 pc, respectively. We considered the situation that the model 
cluster is initially in the extremely subvirial state; the initial virial 
ratio was set to 0.02 which is 25 times smaller than that of the virialized 
stellar system. However, the effects of the stellar evolution 
and gas expulsion affecting the potential of the cluster were 
ignored because there is no clear evidence for supernova 
explosions in this SFR.

We monitored the dynamical evolution of the model cluster by 
taking snapshots at given times. The model cluster undergoes 
collapse for the first 2 Myr and then begins to radially expand. 
The $\Phi$ and tangential velocities ($V_t$) of the simulated 
stars were computed for comparison with those of stars in W4, 
where $V_t$ was calculated from the quadratic sum of 
$V_{\mathrm{R.A.}}$ and $V_{\mathrm{Dec.}}$ after 
subtracting the system velocity. Random errors on PM were 
introduced to the $V_t$ of these stars, based on 
the observational error distributions. In addition, the number 
ratio of low-mass stars in mass bins between 1 and 3$M_{\sun}$ 
from the Kroupa initial mass function \citep{K01} was adjusted 
to that of the observed stars \citep{SBC17} to reproduce the 
incompleteness of our observations.

\begin{figure}[t]
\epsscale{1.0}
\plotone{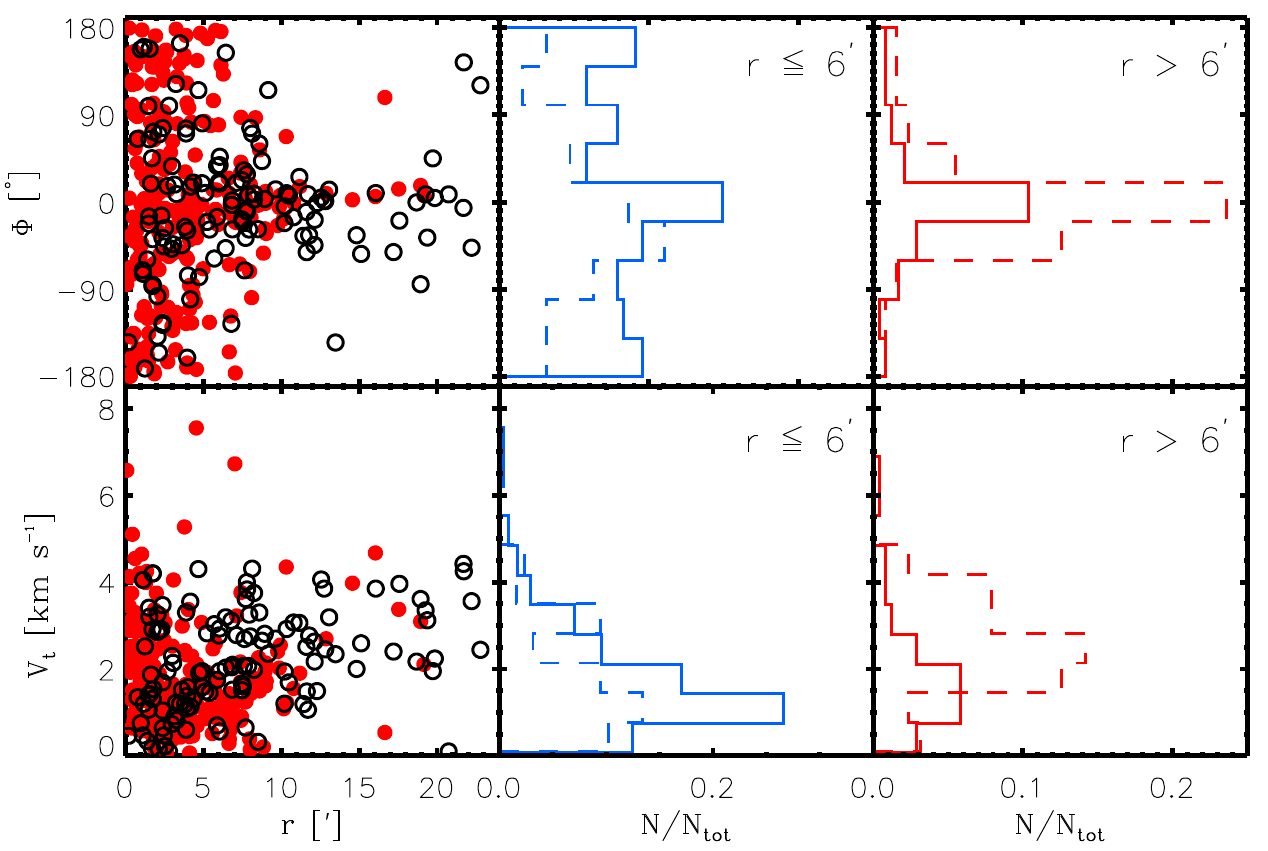}
\caption{Comparison of a model cluster with W4. The radial 
variations of $\Phi$ and $V_t$ obtained from an $N$-body 
simulation (red filled circle) and observation (black open circle) 
are plotted in the upper and lower left-hand panels, respectively. 
The dynamical evolution of a cluster at 3.9 Myr after the initial 
collapse was simulated. The middle and right-hand panels 
display the number distributions of $\Phi$ and $V_t$ for stars 
in the inner ($r \leq 6^{\prime}$) and outer ($r > 6^{\prime}$) 
regions, respectively. The simulated distributions are shown 
by solid lines, while dashed lines represent the observed 
distributions. } \label{fig4}
\end{figure}

Fig.~\ref{fig4} compares the observed radial distributions 
of $\Phi$ and $V_t$ with the simulated results at 3.9 Myr 
after the initial collapse. Note that this timescale does not 
necessarily mean the stellar age (3.5 Myr -- \citealt{SBC17}). 
Our simulation well reproduces the global trend in the radial 
variation of $\Phi$: the isotropic motion in the inner regions 
($r \leq 6^{\prime}$) and the outward motion in the outer regions 
(upper panels of Fig.~\ref{fig4}). This result is also compatible 
with the simulation of early dynamical evolution of young 
star clusters that are dynamically cold and isolated from 
the external tidal field, and the discrepancy around 
$r \sim 20^{\prime}$ between the simulation and 
observation is presumably due to the effect of Galactic 
tide (see figure 4 of \citealt{VVM14}).  

The $V_t$ of stars in the outer region (lower panels of 
Fig.~\ref{fig4}) appear to increase with their radial 
distances in both simulation and observation. This 
radial variation is the consequence of close three-body 
encounters with massive stars during the collapse 
\citep{BKO12,PS12,OK16,GBRT17}. Several 
young runaway stars that presumably originated from 
this dynamical process have been identified 
around the Orion Nebula \citep{MK19}. The $V_t$ 
distribution of simulated stars in the outer region does 
not exactly match the observed one. The number of the 
escaping stars and their velocities can be increased 
or decreased depending on the strength of stellar feedback 
\citep{GBRT17} and the levels of substructures 
\citep{SPA19}. In addition, the latter condition can lead 
to violent dynamical evolution on a very short timescale 
\citep{MVP07,AGP09,AGP10}.

In conclusion, a single star cluster (IC 1805) with or 
without substructures formed in the W4 molecular cloud, 
and then this cluster might have experienced a cold collapse 
in the early epoch. Subsequently, a group of stars escaping 
from the cluster during the expanding phase might form 
the current DSP. Hence, the formation of the structure 
that extends over 20 pc in W4 can be understood in the 
context of dynamical evolution.

On the other hand, several groups of young stars are 
found at the border of the H {\scriptsize \textsc{II}} bubble 
surrounding W4 \citep{PSP19}. The current 
positions of these stars cannot be explained because 
the crossing time is larger than their age. These groups 
are mostly composed of low-mass PMS stars ($<$ 4$M_{\sun}$), 
and they are $\sim2$ Myr younger than the IC 1805 members 
\citep{PSP19}. Therefore, the origin of these young stars 
may be related to feedback-driven star formation, rather than 
dynamical evolution of IC 1805. Indeed, circumstantial 
evidence for feedback-driven star formation has been 
steadily reported in other star-forming regions 
\citep[etc]{FHS02,SHB04,KAGH08,LSK14}. In the case of 
the SFR W8, the fraction of the second generation of stars 
accounts for at least 18\% of the total stellar population 
\citep{LSB18}, which implies that their contribution is far 
from being negligible. Hence, combining two different 
origins (dynamical evolution and feedback driven star formation) 
can help us better understand star formation taking place 
in OB associations. 

\section{SUMMARY}
OB associations are the birth places of young 
stellar population in the Galactic disk, and they 
are ideal sites to understand star formation 
taking place at large spatial scales. W4 in 
the Cas OB6 association is one of active massive 
SFRs in the Galaxy. This SFR is composed of the 
young open cluster IC 1805 and a DSP 
surrounding the cluster. This structural 
feature is probably the relic of the formation process 
of W4. In this work, we investigated the origin 
of this structure using stellar kinematics.

The PMs from the recent {\it Gaia} 
DR2 \citep{gaia18} and RVs measured from 
high-resolution spectra were used to select 
bona-fide members and to probe their velocity 
fields. A total of 127 out of 358 candidates 
were confirmed to be kinematic members of W4. Members in 
the core have an almost isotropic motion, and 
their dynamical state is close to equilibrium. 
On the other hand, members belonging to 
the DSP show a clear pattern of radial expansion.

We considered the early dynamical evolution 
of a star cluster in subvirial state and 
performed an $N$-body simulation. The properties 
of a model cluster were compared with the 
observed ones. Although we did not take into 
account the effects of stellar evolution and gas 
expulsion on the dynamics of the cluster, this 
simulation well reproduced the radial variation of 
projected stellar motions. Hence, 
our results suggest that the origin of the DSP 
distributed over 20 pc is the result of the dynamical 
evolution of IC 1805.

\acknowledgments
The authors thank the anonymous referee for many constructive 
comments and suggestions. B.L. would like to express thanks 
to Professor Gregor Rauw and Dr. Ya\"el Naz\'e for helpful 
discussion, Professor Hwankyung Sung and Professor Mark 
Heyer for providing supplementary data, and ShiAnne Kattner for 
assisting with Hectochelle observations. Observations reported here 
were obtained at the MMT Observatory, a joint facility of the University of Arizona 
and the Smithsonian Institution. In addition, this paper has made use of data 
obtained under the K-GMT Science Program (PID: MMT-2018B-1) 
funded through Korean GMT Project operated by Korea Astronomy 
and Space Science Institute (KASI) and from the European Space Agency (ESA) 
mission{\it Gaia} (https://www.cosmos.esa.int/gaia), processed by the {\it Gaia}
Data Processing and Analysis Consortium 
(DPAC, https://www.cosmos.esa.int/web
/gaia/dpac/consortium). Funding for the DPAC has been 
provided by national institutions, in particular the institutions 
participating in the {\it Gaia} Multilateral Agreement. This work 
was supported by the National Research Foundation of Korea (NRF) 
grant funded by the Korea government (MSIT) (Grant No: 
NRF-2019R1C1C1005224). N.H. acknowledges 
support from the Large Optical Telescope Project operated by 
KASI. B.-G.P. acknowledges support from the K-GMT Project 
operated by KASI. 

% J.-E.L. acknowledges support from the Basic Science 
%Research Program through the NRF (grant No. NRF-2018R1A2B6003423) 
%and the KASI under the R\&D program supervised by the Ministry 
%of Science, ICT and Future Planning. 

{\it Facilities:} \facility{MMT (Hectochelle)}.

\end{document}